\documentstyle[12pt]{article}
\input psfig
 1
\def\citenum#1{{\def\@cite##1##2{##1}\cite{#1}}}
\def\citea#1{\@cite{#1}{}}
\def\l{\lambda}

\def\({\left(}
\def\){\right)}

\def\citenum#1{{\def\@cite##1##2{##1}\cite{#1}}}
\def\citea#1{\@cite{#1}{}}

\def\l1vt{\vec{l_{1\perp}}}

\def\rt{r_{\perp}}
\def\bt{b_{\perp}}
\def\rt2{$r^2_{\perp}$}
\def\bt2{$b^2_t$}

\def\jol1{$J_0(\,l_{1\perp}\,r_{\perp}\,)$}

\def\citea#1{\@cite{#1}{}}

\def\beq{\begin{equation}}
\def\eeq{\end{equation}}
\def\bea{\begin{eqnarray}}
\def\eea{\end{eqnarray}}

\def\bbbz{{\mathchoice {\hbox{$\sf\textstyle Z\kern-0.4em Z$}}
{\hbox{$\sf\textstyle Z\kern-0.4em Z$}}
{\hbox{$\sf\scriptstyle Z\kern-0.3em Z$}}
{\hbox{$\sf\scriptscriptstyle Z\kern-0.2em Z$}}}}
\relax
\begin{document}
\begin{titlepage}
\noindent
 June 16 1995   \hfill  CBPF-NF-010/95 \,\,\,\, {\bf hep - ph / 9503399}\\[4ex]
\begin{center}
{\Large\bf{GLUON  STRUCTURE FUNCTION}\\[1.4ex]
{\Large \bf  FOR DEEPLY INELASTIC SCATTERING}\\[1.4ex]
{\Large\bf WITH NUCLEUS IN QCD}  \\[9ex]
{\large M.\,\, B.\,\, G A Y \,\,D U C A T I  and  A L V A R O  \,\, L.\,\,
A Y A L A\,\,  F$^{\underline{o}}$}}\\[1.5ex]
{\it Instituto de Fisica, Univ. Federal do Rio Grande do Sul}\\
{\it Caixa Postal 15051, 91500 Porto Alegre, RS, BRASIL}\\[1.5ex]
{\large and}\\[1.5ex]
{\large E U G E N E \,\, L E V I N}   $^{\dagger)} $
\footnotetext{$^{\dagger})$ Email: levin@lafex.cbpf.br; levin@fnalv.fnal.gov;
levin@ccsg.tau.ac.il} \\[1.5ex]
{\it  LAFEX, Centro Brasileiro de Pesquisas F\'\i sicas  (CNPq)}\\
{\it Rua Dr. Xavier Sigaud 150, 22290 - 180 Rio de Janeiro, RJ, BRASIL}
\\{\it and}\\
{\it Theory Department, Petersburg Nuclear Physics Institute}\\
{\it 188350, Gatchina, St. Petersburg, RUSSIA}\\[6.5ex]
\end{center}
{\large \bf Abstract:}
In this talk we present the first calculation of the gluon
structure function for nucleus in QCD. We discuss the Glauber formula for the
gluon structure function and the violation of this simple approach
 that we anticipate in QCD
\footnote{Talk given by E. Levin at QCD and nuclear target session
at the Workshop on Deep Inelastic Scattering and QCD,
Paris, April 1995}.
\end{titlepage}
\section{Introduction.}
The subject of the talk is the gluon structure function for DIS with nucleus.
The gluon structure function is the most important physical observable that
governs the physics at high energy (low Bjorken $x$) in the DIS. Dealing
with nucleus we have to take into account the shadowing correction, which is
the
main point of interest in this talk. We show that the shadowing correction in
the region of small $x$ can be treated
theoretically in QCD and can be reliably calculated using
the information on the behaviour of the gluon structure function for the
nucleon. We organize the presentation in the following way: first, we
discuss the Glauber approach to the nucleus gluon structure function and
answer the question what information on nucleon structure function we need
to provide a reliable calculation using the Glauber formula; second, we
briefly consider the corrections to Glauber approach that have been
anticipated in QCD. It should be stressed that this is the first
presentation of
our results and the lack of space does not allow us to discuss the issue in
details. This is why we are going to outline our strategy and to present
the first estimates rather than to give the complete study of the problem
which will be published elsewhere.

\section{Glauber approach in QCD .}
The idea how to write the Glauber formula in QCD has been first formulated in
ref. \cite{LERY} and was carefully developed by Mueller in ref. \cite{M90}.
It is easier to explain the idea considering the penetration of quark-antiquark
pair through the target.
Indeed, during the time of passage through the target the transverse
distance $r_t$
between quark and antiquark can vary by the amount $ \Delta r_t\,\,\propto
\,\,R
\,\,\frac{k_t}{E}$, where  $E$ is
the energy of the pair and $R$ is the size of the target (see Fig.1).
The quark transverse momentum ($k_t$) is $k_t\,\,\propto\,\frac{1}{r_t}$ due to
uncertainty principle. Therefore
\begin{equation}
\Delta \,r_t\,\,R\,\,\frac{k_t}{E}\,\,\ll\,\,r_t
\end{equation}
holds if
\begin{equation}
\,r^2_t\,\cdot\,s\,\,\gg\,\,\,2\,m\,R
\end{equation}
In terms of Bjorken $x$ the above condition looks as follows:
\begin{equation}
x\,\,\ll\,\,\frac{1}{2\,m\,R}
\end{equation}
It means that the transverse distance between quark and antiquark is a  good
degree of freedom \cite{LERY}\cite{M90}\cite{MU}. As has been shown by Mueller
not only quark - antiquark pair can be considered in such way. The propagation
of a gluon through the target can be treated in a similar
 way as the interaction of gluon - gluon pair with definite transverse
 separation $r_t$ with the target. The total cross section of the absorbtion
of gluon($G^*$) with virtuality $Q^2$ and Bjorken $x$ can be written in the
form:
\begin{equation}
\sigma_{G^*}\,\,=\,\,
\end{equation}
$$
\int^1_0 d z \,\,\int \,\frac{d^2 r_t}{2 \pi}\,\,
\int\frac{d^2 b_t}{2 \pi}
\Psi^{G^*}_{\perp} (Q^2, r_t,x,z)\,\,
2\,\cdot\,
\{1\,\,-\,\,exp[ -\,\,\sigma(r^2_t)\,\,
S(b^2_t)\, ]\,\}\,\cdot\,{\Psi^{G^*}_{\perp}}^* (Q^2, r_t,x,z)
$$
where $\Psi^{G^*}_{\perp}$ is the wave function of the virtual gluon with
transverse polarization.
%\ffig{fig1paris.eps}{40mm}{\em The structure of the parton cascade in the
% Glauber formula }{fig1}
\begin{figure}[htbp]
\centerline{\psfig{figure=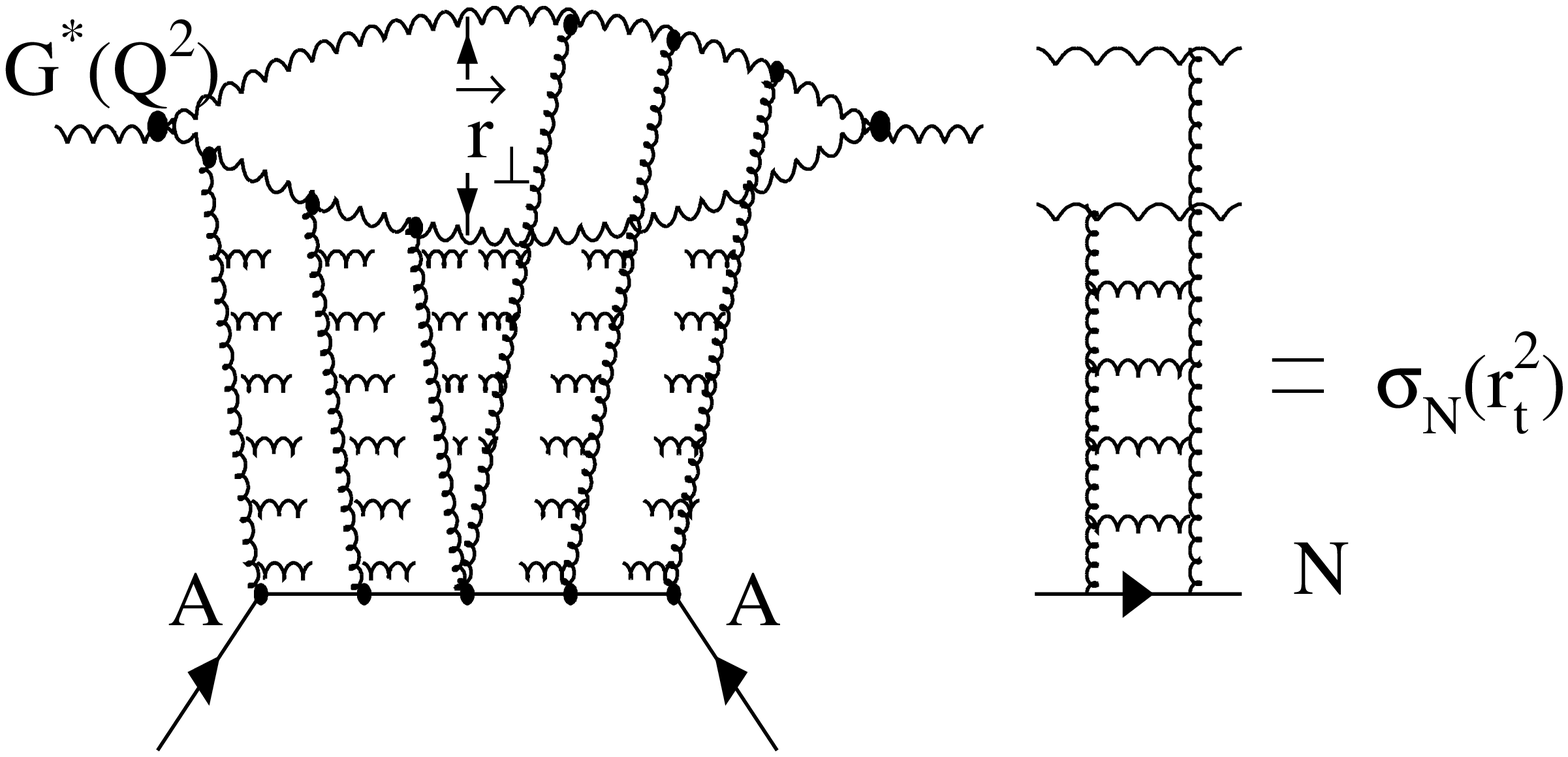,width=120mm,height=120mm}}
\caption{\em The structure of the parton cascade in the Glauber formula. }
\label{Fig.1}
\end{figure}
 As was shown in ref. \cite{M90} within leading log
approximation of perturbative QCD (pQCD) we can safely replace this function
 by $\frac{1}{r^2_t}$ after integration over $z$ in eq.(3). Finally the Glauber
formula for the gluon structure function reads ( for $N_c = N_f$ = 3):
\begin{equation}
x G(Q^2,x)\,\,=\,\,\frac{4}{ \pi^2}\,\,\int^1_x \,\frac{d x'}{x'}
\int\,\,
\frac{d^2 \,b_{t}}{\pi}\int^{\infty}_{\frac{4}{Q^2}}
\,\frac{d^2\,r_{t}}{\pi}\,\,
\frac{1}{r^4_{t}}
\,\,2\,\{1\,\,-\,\,e^{-\frac{1}{2}\,\sigma^{GG}(r^2_{t},x')
\,S(b^2_{t})}\}
\end{equation}
where
\begin{equation}
\sigma^{GG}\,\,=\,\,\frac{3 \alpha_s}{4}\,\,\pi^2\,\,
r^2_{t}\,\,x G(\frac{4}{r^2_{t}},x)
\end{equation}
and $S(b^2_t)$  is the profile function in impact parameter space for
the interaction of the gluon-gluon pair with the target. We use for the
 calculation the Gaussian parameterization for $S$, namely:
\begin{equation}
S(b^2_t)\,\,=\,\,\frac{A}{\pi R^2_A}\,\,e^{-\,\,\frac{b^2_t}{R^2_A}}
\end{equation}
where $A$ is the number of the nucleons in a nucleus and
 $R^2_A$ is the mean radius of a nucleus, which is equal to
$$
R^2_A\,\,=\,\,\frac{2}{5}\,\,R^2_{WS}
$$
$R_{WS}$ is the size of the nucleus in the Wood - Saxon parameterization, which
we chose $R_{WS}\,\,=\,\,r_0 A^{\frac{1}{3}}$ with $r_0 = 1.3 fm$.
Using the Gaussian parameterization for $S$ we can take the integral over $b_t$
and get the answer:
\begin{equation}
x G(Q^2,x)\,\,=\,\,\frac{2}{ \pi^2}\,\,\int^1_x \,\frac{d x'}{x'}
\,\,\int^{\infty}_{\frac{1}{Q^2}}
\,\frac{d^2\,r'_{t}}{\pi}\,\,
\frac{R^2_A}{r'^4_{t}}
\,\,\{ \,C\,\,+\,\,\ln \kappa_G(x',r'^2_t)\,\,+\,\,E_1 (
\kappa_G(x',r'^2_t)\,\,
\}
\end{equation}
where $C$ is the Euler constant and $E_1$ is the exponential integral
 (see ref.\cite{AB} {\bf 5.1.11}) and
\begin{equation}
\kappa_G(x',r'^2_t)\,\,=\,\,\frac{3 \alpha_s A \pi \,\,r'^2_t}{2  R^2_A}\,\,
x' G_N(x',\frac{1}{r'^2_t})
\end{equation}
\section{Theory status of the Glauber formula.}
In this section we would like to recall the main assumptions that have been
made to get the Glauber formula:

1.Energy ($x$) should be so high  (small)  to satisfy eqs.(2) and (3) and
$\alpha_s \ln (1/x)\,\,\approx\,\,1$. The last condition means that we are
doing
the calculation in leading log(1/$x$) approximation of perturbative QCD (pQCD).

2.The GLAP \cite{GLAP} evolution equation holds in the region of small $x$.
It means that $\alpha_s \ln (1/ r^2_t) \,\,\approx \,\,1$. One of the lessons
that we have learned at this workshop is the fact that the GLAP equation
 is able to describe the HERA data quite well.

3.Only the fastest partons ( $GG$ pair) interacts with the target and there are
no correlations between partons from different parton cascades (see Fig.1).

4.There are no correlations between different nucleons in a nucleus.

5.The average $b_t$ for $GG$ pair - nucleon interaction is much smaller than
$R_A$.

We are going to discuss how well all the above assumptions work in the last
 section of the talk.
\section{Results.}
In our calculations we use the GRV parameterization \cite{GRV} for the nucleon
gluon structure function. This parameterization describes the data quite well
and it is suited for our purpose because (i) the initial virtuality for
the GLAP equation is small and we can discuss the contribution of the large
distances having some support in the experimental data; (ii) the
parameterization
uses the GLAP equation and the most essential contribution comes from the
region where $\alpha_s \ln Q^2 \,\,\approx\,\,1$ and $\alpha_s \,\ln (1/x)
\,\,\approx\,\,1$.
\subsection{Where the shadowing corrections are big.}
Fig.2 shows the kinematic region of the deeply inelastic scattering. The
curves are the solution of the equation $\kappa_G\,\,=\,\,1$ for N (nucleon),
Ca and Au.
Above each of these curves the value of $\kappa_G\,>\,1$ and the shadowing
correction (SC) are big, below $\kappa_G\,<\,1$ and the SC
are rather small.
\begin{figure}[htbp]
\centerline{\psfig{figure=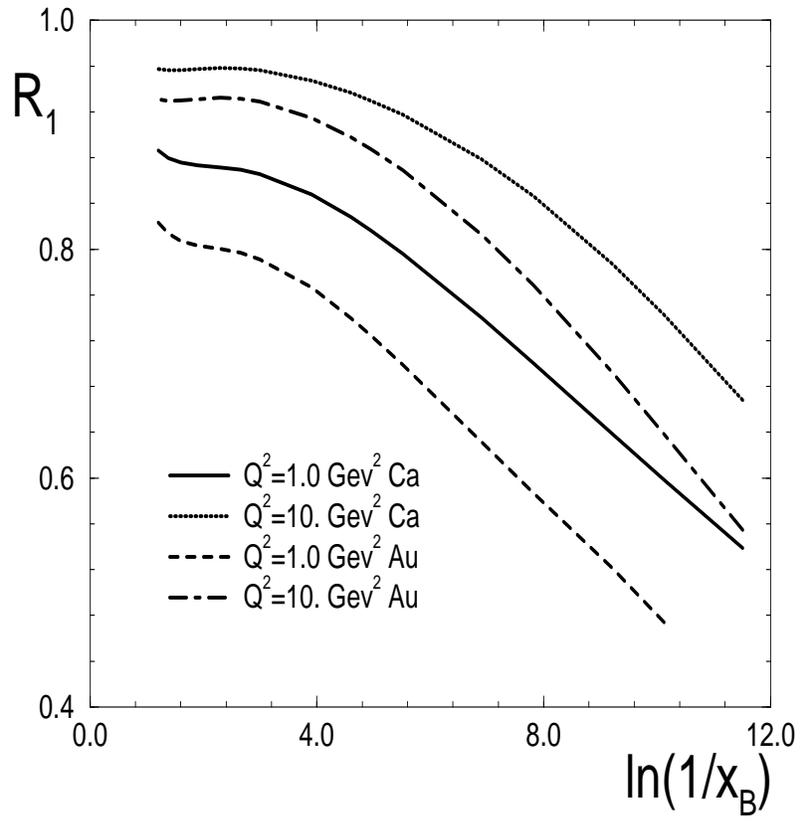,width=120mm,height=120mm}}
\caption{\em Solution for $\kappa=1$. }
\label{Fig.2}
\end{figure}

\subsection{What we are able to calculate in QCD.}
{}From the master equation (5) one can see that the large distances contribute
 to the  value of the gluon structure function. Such contributions we
 are not able to calculate in pQCD and the value of the gluon structure
function crucially depends on the hypothesis about nonperturbative behaviour
of the gluon structure function that we have to assume to treat the
 large distances contribution. In pQCD we can safely calculate only the
difference $x G_A(x,Q^2)\,\,-\,\,xG_A(x,Q^2=Q^2_0)$ where $Q^2_0$ is the
initial virtuality. In Fig. 3 one can find the calculation for the ratio:
\begin{equation}
R_1\,\,=\,\,\frac{x G_A(x,Q^2)\,\,-\,\,xG_A(x, Q^2 = Q^2_0)}{A\,(\,
x G_N(x,Q^2)\,\,-\,\,x G_N (x,Q^2 = Q^2_0))}
\end{equation}
 as function of $x$ for Ca and Au ($Q^2_0 $ = 0.25 $GeV^2$).
\begin{figure}[htbp]
\centerline{\psfig{figure=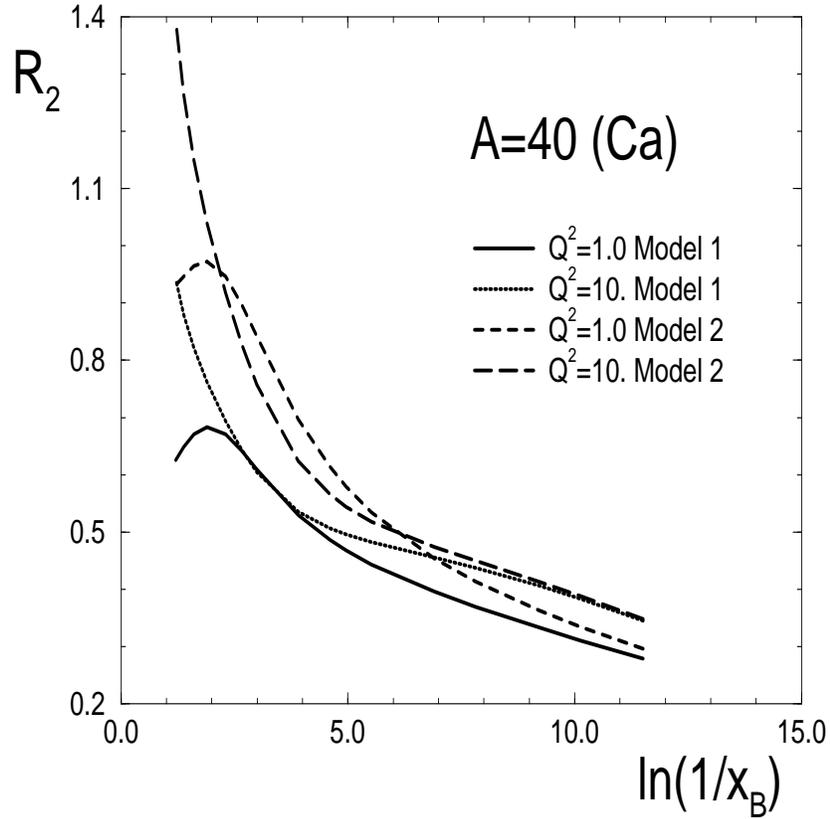,width=120mm,height=120mm}}
\caption{\em $R_{1}$ as a function of $ln(1/x_{B})$ for $Ca$ and $Au$.}
\label{Fig.3}
\end{figure}

\subsection{Contribution of the large distances.}
As has been mentioned we are able to treat this problem only using some model
for large distance behaviour of $xG_N$.  Fig.4 shows the ratio:
\begin{equation}
R_2\,\,=\,\,\frac{x G_A(x,Q^2)}{A\,(\,
x G_N(x,Q^2))}
\end{equation}
for two models:

1. $x G_N(x, Q^2 \,<\,Q^2_0)\,\,=\,\,\frac{Q^2}{Q^2_0} \,x G_N(x,Q^2=Q^2_0)$.
This model takes into account the correct limit of the gluon structure
function at small value of $Q^2$, which follows from the gauge invariance of
QCD.

2.$x G_N(x, Q^2 \,<\,Q^2_0)\,\,=\,\,\,x G_N(x,Q^2=Q^2_0)$.
In this model we assume that the scale for the behaviour $xG_N \propto Q^2$
is much smaller than $Q^2_0$.
\begin{figure}[htbp]
\centerline{\psfig{figure=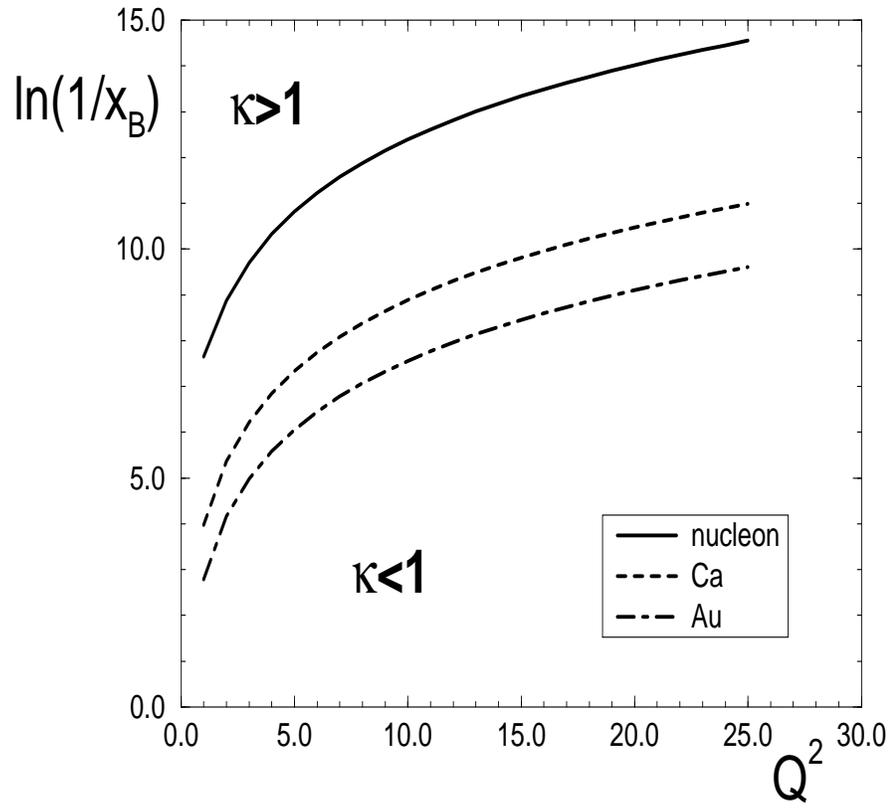,width=120mm,height=120mm}}
\caption{\em $R_{2}$ ratio for two models. }
\label{Fig.4}
\end{figure}
We look at the difference in the value of $R_2$ as the estimates
of possible errors that originated from our poor knowledge of long distance
behaviour of the gluon structure function. The conclusion is that we
cannot calculate the value of $R_2$ even at $x = 10^{-3}$ at $Q^2 = 1GeV^2$
with better accuracy that 20\%, while at larger value of $Q^2$
 ($Q^2 \sim 10 GeV^2$) the accuracy is better ( about 5\%).
\section{Correction to the Glauber formula.}
To abandon the main assumptions which have been made in the Glauber
formula we have to develop a technique to include (i) the interactions of
all partons ( not only the fastest one) with a nucleus; (ii) the parton
interaction inside a nucleon and (iii) the nucleon correlation inside a
nucleus.
 Such a technique has been suggested in ref.\cite{LALE} and it is
based on new evolution equation that takes into account the parton interaction
inside the parton cascade as well as the parton interaction with the different
nucleons. The lack of space does not allow us to discuss this problem in
 details but we want to point out that the Glauber formula shall be used
as initial condition to the new evolution equation of ref.\cite{LALE}.

The numerical estimates \cite{LALE}
 shows that the most essential contribution at least in
HERA kinematic region is generated by the interaction of all partons with the
target which corresponds to so called ``fan" diagrams (see ref.\cite{GLR})
while the dynamic correlation (see refs.\cite{LALEREV} \cite{HT})
due to the interaction inside the parton cascade
remains to be rather small.

 However, reliable estimates can be done only
after extracting from HERA data the value of the scale for the SC for nucleon
structure function. In our numerical estimates which will be published
elsewhere we follow the following strategy: we neglect the dynamic
 gluon correlations and iterate the master equation (5) several times.
Since there is strong ordering in rapidity of parton
in each parton cascade the $``i-th" $ iteration means
 that we take into account the interaction with the target of all partons
with the value of rapidity ($y$) larger than $y_i$. It turns out that we
have to make only two iterations for $x > 10^{-3}$ to get convergent result.
\section{Conclusions.}
We know the Glauber formula and the technique how to find the corrections to
the Glauber formula in QCD. However we cannot provide reliable
predictions for the gluon structure function for nucleus until we will
get more data on low $Q^2$ and low $x$ behaviour of the nucleon structure
 function. Unfortunately, we have to know not only the behaviour of the
nucleon structure function but also extract from the experimental data
the scale for the shadowing corrections to the value of the gluon
structure function in the nucleon. We are going to check how the
 information from DIS with nucleus can reduce these uncertainties.
At the moment we suggest to measure the ratio $R_1$ which can be calculated
with much better accuracy than the value of the gluon structure function.

\begin{center}
{\large\bf Acknowledgements}
\end{center}
It is a pleasure to thank the convenors of the
 QCD  and nuclear target working group for giving us time to discuss our
results
and the organizing committee of
the conference for the working atmosphere  at the conference. Work partially
financed by CAPES and CNPq, Brazil.

\end{document}